\documentclass[twocolumn,showpacs,preprintnumbers,amsmath,amssymb]{revtex4-1}
\usepackage{graphicx}
\usepackage{rotating}
\begin{document}

\renewcommand{\theequation}{\thesection.\arabic{equation}}

\newcommand{\re}{\mathop{\mathrm{Re}}}

\newcommand{\be}{\begin{equation}}
\newcommand{\ee}{\end{equation}}
\newcommand{\bea}{\begin{eqnarray}}
\newcommand{\eea}{\end{eqnarray}}

\title{Cosmological tests of sudden future singularities.}

\author{Tomasz Denkiewicz}
\email{atomekd@wmf.univ.szczecin.pl}
\author{Mariusz P. D\c{a}browski}
\email{mpdabfz@wmf.univ.szczecin.pl}
\affiliation{\it Institute of Physics, University of Szczecin, Wielkopolska 15, 70-451 Szczecin, Poland}
\affiliation{\it Copernicus Center for Interdisciplinary Studies,
S{\l }awkowska 17, 31-016 Krak\'ow, Poland}

\author{Hoda Ghodsi}
\email{h.ghodsi@ipm.ir}
\affiliation{\it School of Physics, Institute for Research in Fundamental Sciences (IPM) Tehran, 
PO. Box 19395-5531, Iran}
\author{Martin A. Hendry}
\email{martin@astro.gla.ac.uk}
\affiliation{\it SUPA, School of Physics and Astronomy,
University of Glasgow, Glasgow G12 8QQ, UK}

\date{\today}

\input epsf

\begin{abstract}
We discuss combined constraints, coming from the cosmic microwave background shift parameter $\mathcal{R}$, baryon acoustic oscillations (BAO) distance parameter $\mathcal{A}$, and from the latest type Ia supernovae data, imposed on cosmological models which allow sudden future singularities of pressure. We show that due to their weakness such sudden singularities may happen in the very near future and that at present they can mimic standard dark energy models.

\end{abstract}


\maketitle

\section{Introduction}
\label{intro}
\setcounter{equation}{0}

Sudden future singularities of pressure \cite{SFS} (SFS or type II singularities \cite{nojiri,aps10}) are some exotic-type singularities which are allowed to occur in the universe. The inspiration for investigating these new types of singularity was the observation of high-redshift Type Ia supernovae (SNIa) which over the past decade has provided strong evidence that the expansion of the universe is accelerating \cite{supernovaeold} and, in particular, the further extension of these datasets \cite{supernovaenew,kowalski} which remain consistent with the dark energy existing in the form of phantom \cite{phantom} energy. Phantom-driven dark energy leads to a big-rip singularity (BR or type I according to \cite{nojiri}) in which all the matter is dissociated by the phantom-driven dark energy which acts as antigravity in a large and a dense universe \cite{caldwellPRL}.
This behavior is of course different from the standard picture of cosmic evolution which allows big-bang (BB) or big-crunch (BC) standard types of singularities only. Besides, phantom energy violates all the energy conditions: the null ($\varrho c^2 + p \geq 0$), weak ($\varrho c^2 \geq 0$ and $\varrho c^2 + p \geq 0$), strong ($\varrho c^2 + p \geq 0$ and $\varrho c^2 + 3p \geq 0$), and dominant energy ($\varrho c^2 \geq 0$, $-\varrho c^2 \leq p \leq \varrho c^2$) (here $c$ is the speed of light, $\varrho$ - the mass density in kg m$^{-3}$, and $p$ - the pressure).  Apart from big-rip and sudden future singularities there are numerous other types such as: generalized sudden future singularities (GSFS), finite scale factor singularities (FSF or type III) which were tested  against observations in \cite{fsf},  big-separation singularities (BS or type IV) and $w$-singularities (type V) \cite{wsing,type0V}. The singularities which fall outside this classification (with perhaps a big-bang as type 0 \cite{type0V}) are curvature singularities with respect to a parallelly propagated basis (p.p. curvature singularites) which show up as directional singularities \cite{LFJ2007} and also intensively studied recently: the little-rip singularities \cite{LRip} and the pseudo-rip singularities \cite{PRip}.  They are characterized by violation of all, some or none of the energy conditions which results in a blow-up of all or some of the appropriate physical quantities such as: the scale factor, the energy density, the pressure, and the barotropic index (for a review see Ref. \cite{aps10}).  In fact, all of these singularities may be investigated by using the higher-order characteristics of the expansion of the universe known as statefinders \cite{statef}.

Sudden future singularity models result in leaving no restrictions on an  equation of state $p=p(\varrho)$ of the cosmological matter,
which allows the unconstrained evolution of the energy density and pressure. The nature of a sudden future singularity is different from that of a standard big-bang singularity and also from an exotic big-rip singularity in that it does not exhibit geodesic incompleteness and the cosmic evolution may eventually be extended beyond it \cite{lazkoz,adam}. The only physical characteristic of these singularities is a momentarily infinite peak of the tidal forces in the universe. In generalized sudden future singularity (GSFS) models this peak may also appear in the derivatives of the tidal forces \cite{SFS}. GSFS, BS, and w-singularities, like SFS, admit geodesic completeness and so they are weak singularities which can be passed through by both point particles and extended objects \cite{LFJ2010, adam}. It is interesting to note that sudden future singularities are in a way similar to yet another type, which were termed finite density singularities \cite{dabrowski93}. The crucial difference is that finite density singularities occur as singularities in space rather than in time, which means that even at the present moment of cosmic evolution they could exist somewhere in the Universe \cite{inhpress}. We will not discuss in detail these finite density singularities in this paper since they basically appear in cosmological models without homogeneity. On the other hand, it is worth mentioning that the sudden future singularities are quite generic, since they may arise in both homogeneous \cite{SFS} and inhomogeneous \cite{sfs1} models of the universe.

In Ref. \cite{PRD07} we discussed the constraints imposed on SFS models which came from SNIa data. We showed that those data were consistent with the occurrence of a sudden future singularity in the very near future -- as little as 8.7 million years from now. In Ref. \cite{ghodsi11} we extended our investigation to confront SFS models with other cosmological data, from the cosmic microwave background (CMB) and baryon acoustic oscillations.  We demonstrated that the class of SFS models introduced in \cite{SFS} was not compatible with current observations, in the particular case where the asymptotic behaviour of the scale factor close to the BB singularity mimics a dust-filled Einstein de Sitter universe.  In this paper we extend our investigation further, confronting current cosmological data with SFS models that simulate the behaviour of more general flat, barotropic fluid models. The paper is organized as follows. In Section \ref{models} we present sudden future singularities in a framework which is appropriate to further discussion. In Sections \ref{supernovae}-\ref{BAO} we discuss current observational data on supernovae, shift parameter and baryon acoustic oscillations in the universe. In Section \ref{conclusion} we compare these observational data with our SFS models and use them to determine constraints on the model parameters and present our conclusions.

\section{The models}
\label{models}

In order to obtain a sudden future singularity consider the simple framework of an Einstein-Friedmann cosmology governed by the standard field equations
\bea \label{rho} \varrho(t) &=& \frac{3}{8\pi G}
\left(\frac{\dot{a}^2}{a^2} + \frac{kc^2}{a^2}
\right)~,\\
\label{p} p(t) &=& - \frac{c^2}{8\pi G} \left(2 \frac{\ddot{a}}{a} + \frac{\dot{a}^2}{a^2} + \frac{kc^2}{a^2} \right)~,
\eea
where the energy-momentum conservation law
\be
\label{conser}
\dot{\varrho}(t) = - 3 \frac{\dot{a}}{a}
\left(\varrho(t) + \frac{p(t)}{c^2} \right)~,
\ee
is trivially fulfilled due to the Bianchi identity. Here $a \equiv a(t)$ is the scale factor, the dot means the derivative with respect to time $t$, $G$ is the gravitational constant, and the curvature index $k=0, \pm 1$. What is crucial to obtain a sudden future singularity is that in the general case no link between the energy density and pressure (i.e. the equation of state) is specified (but see below for the case of the flat models).

From equations (\ref{rho})-(\ref{p}) one can easily see that a pressure singularity $p \to \infty$ occurs when the acceleration $\ddot{a} \to - \infty$, no matter that the value of the energy density $\varrho$ and the scale factor $a(t)$ are regular. Since in that case $\mid p \mid > \varrho$, it is clear that the dominant energy condition is violated.
This condition can be achieved if the scale factor takes the form \cite{SFS,BGT,PRD07}
\be \label{sf2} a(y) = a_s \left[\delta + \left(1 - \delta \right) y^m -
\delta \left( 1 - y \right)^n \right]~, \hspace{0.5cm} y \equiv t / t_s \ee
with the appropriate choice of the constants $\delta, t_s, a_s, m, n$. First publication of the eq.(\ref{sf2}) was in \cite{BGT} where it was used to show that closed universes obeying energy conditions did not necessarily recollapse.
Notice that in equation (\ref{sf2}) $a_s$ has the unit of length and all the terms (including $y$) in the bracket are dimensionless.
It is interesting to note that for a flat ($k=0$) Friedmann model we have an explicit relation between the pressure and the energy density, though with a time-dependent barotropic index, in the form
\be
p_s(t) = w_s(t) \varrho_s(t)~~,
\label{pete}
\ee
where
\be
w_s(t) = \frac{c^2}{3} \left[2q(t) - 1 \right]~~,
\label{wute}
\ee
and $q(t) = - \ddot{a}a/\dot{a}^2$ is the deceleration parameter. The index `$s$' has been attached to mark the fact that we have a type of matter which is of sudden future singularity origin and this may play the role of the dark energy. In the limit $\delta \to 0$ we have $q(t) = (1-m)/m$ which for dust (i.e. for $m=2/3$) gives $q(t) = 1/2$, as required. In fact, the relations (\ref{pete}) and (\ref{wute}) remain of the same form if the time $t$ is replaced by the scaled time to a sudden singularity $y$.

The standard Friedmann limit (i.e. models without an SFS) of (\ref{sf2}) is achieved when $\delta \to 0$; hence $\delta$ becomes the ``non-standardicity" parameter of SFS models. Additionally, notwithstanding Ref. \cite{SFS} and in agreement with the field equations (\ref{rho})-({\ref{p}), we assume that $\delta$ can be both positive and negative leading to a deceleration or an acceleration (cf. (\ref{sf2})) of the universe, respectively.

It is important to our discussion that the asymptotic behaviour of the scale factor (\ref{sf2}) close to the BB singularity at $t=0$ is given by a simple power-law $a_{\rm BB} = y^m$, simulating the behaviour of flat $k=0$ barotropic fluid models with $m = 2/[3(w+1)]$~. This allows us to preserve all the standard observed characteristics of early universe cosmology -- such as the cosmic microwave background, density perturbations, nucleosynthesis etc. -- provided we choose an appropriate value of $m$. On the other hand, close to an SFS the asymptotic behaviour of the scale factor is non-standard, $a_{\rm SFS}(y) = a_s \left[ 1 - m \left( 1 - \delta \right) \left(1 - y\right)~ \right]$, showing that $a_{SFS}=a_s$ for $t=t_s$ (i.e. $y=1$) at the SFS. Notice that one does not violate the energy conditions if the parameter $m$ lies in the range \be \label{m} 0 < m \leq 1~ \hspace{0.5cm} (w \geq -1/3), \ee
This range of values is, in fact, equivalent to a standard (neither quintessence-like nor phantom-like) evolution of the universe. However, with no adverse impact on the field equations (\ref{rho})-(\ref{p}), one could also extend the values of $m$ to lie in the complementary ranges \cite{lazkoz} $m > 1$ (i.e. $-1<w<-1/3$) for quintessence, and  $m < 0$ (i.e. $w<-1$) for phantom models, although these ranges may lead to violation of the strong and weak energy conditions, respectively.

A special case in which sudden future singularities allow for an explicit equation of state is the anti-Chaplygin gas model \cite{kamenschchik}
\be
p(t) = \frac{A}{\varrho(t)} \hspace{0.5cm} (A \geq 0)~~,
\label{eoschap}
\ee
and the pressure singularity, known in this case as the so-called `Big-Brake', is achieved for $\varrho \to 0$ and so $p \to \infty$. Here we can easily check that one is able to obtain an anti-Chaplygin gas behavior for the scale factor (\ref{sf2}). In order to prove that we need simply to consider the first time derivative of (\ref{sf2})
\be
\dot{a}(t) = a_s \left[\frac{m(1 - \delta)}{t_s} y^{m-1} + \delta \frac{n}{t_s} \left(1-y \right)^{n-1} \right]~.
\ee
We require $\dot{a} \to 0$, which corresponds to $\varrho \to 0$ at $y=1$ so that we have a condition that either $m \to 0$ or $\delta \to 1$. In fact, these conditions are almost equivalent since
\bea
\lim_{m \to 0} a(y) &=& a_s [1 - \delta (1-y)^n ]~,
\label{achap1}\\
\lim_{\delta \to 1} a(y) &=& a_s [1 - (1-y)^n ]~.
\label{achap2}
\eea
It follows that (\ref{achap1}) is more general since additionally it does not restrict $\delta$l, and moreover it has a standard Friedmann limit $\delta \to 0$, though a static one. It is useful to calculate from (\ref{achap1}) that (cf. (\ref{Hy}))
\bea
H(y) &=& \frac{n\delta(1-y)^{n-1}}{[1 - \delta(1-y)^n]}~,\\
q(y) &=& \frac{(n-1)}{n\delta} \left[\frac{1}{(1-y)^n} - \delta \right]~.
\eea

Using the conservation law (\ref{conser}) together with the barotropic relation (\ref{pete}) and (\ref{wute}) (which holds for flat $k=0$ models only) we may write
\be
\frac{d\varrho_s}{\varrho_s} = - 2 \left[q(t) + 1 \right] \frac{\dot{a}}{a} dt ~,
\ee
which integrates to give
\bea
\varrho_s(t) &=& \varrho_{0s} \exp{\left[-2 \int_{t_0}^{t} \left[q(t') + 1 \right] H(t') dt'\right]}
\label{roqu}
\\
&=& \varrho_{0s} \exp{\left[-3 \int_{t_0}^{t} \left[w(t') + 1 \right] H(t') dt'\right]}~.
\label{rowu}
\eea

The following regimes of the relations (\ref{roqu}) and (\ref{rowu}) are in order. Firstly, the regime of the early universe can be recovered by taking $t \ll t_s$ (which is equivalent to $\delta \to 0$) one has
\be
\varrho_s/\varrho_{0s} = (y_0/y)^2 \propto t^{-2}~.
\ee
Secondly, the near-to-SFS limit can be obtained by taking $t \simeq t_s$ which gives
\be
\varrho_s(t) = \varrho_{0s} \left[\frac{(1-\alpha)t_s + \alpha t}
{(1-\alpha)t_s + \alpha t_0} \right]^2 \propto \varrho_{0s} t^2~,
\label{rosfs}
\ee
with $\alpha = m(1 - \delta)$. In fact, one can see from (\ref{rosfs}) that the contribution from the `SFS-driven' matter is more and more important and starts dominating from the current moment of the evolution, i.e., $t < t_0$ implies $\varrho_s < \varrho_{s0}$, $t = t_0$ implies $\varrho_s = \varrho_{0s}$,
$t > t_0$ implies $\varrho_s > \varrho_{s0}$, and $t \to t_s$
implies
$\varrho_s = \varrho_{0s} t_s^2 \left(t_s - \alpha t_s + \alpha t_0 \right)^{-2}$.

Defining the density parameter
\be
\Omega_{s0} = \frac{8\pi G}{3 H_0^2} \varrho_{s0}~,
\ee
one can write down the flat Friedmann equation (\ref{rho}) in the form
\bea
H^2(t) &=& \frac{\dot{a}^2}{a^2} = \frac{8\pi G}{3} \varrho_s(t)
 \\
&=& H_0^2 \Omega_{s0} \exp{\left[-3 \int_{t_0}^{t} \left[w(t') +1 \right] H(t') dt' \right]}~, \nonumber
\eea
or
\be
H(y) = H_0 \sqrt{\Omega_{s0} \exp{\left[-3 \int_{y_0}^{y} \left[w(y') + 1 \right] H(y') dy' \right]}}~.
\ee

In this paper we consider three observational constraints on SFS models: luminosity distance moduli to SNIa, the CMB shift parameter, which is a scaled distance to the last scattering surface of the cosmic microwave background, and the distance parameter constrained by baryon acoustic oscillations.

\section{Supernovae}
\setcounter{equation}{0}
\label{supernovae}

We proceed within the framework of Friedmann cosmology, and consider an observer located at $r=0$ at coordinate time $t=t_0$. The observer receives a light ray emitted at $r=r_1$ at coordinate time
$t=t_1$ and, according to (\ref{sf2}), with redshift given by
\be
\label{redshift}
1+z=\frac{a(t_0)}{a(t_1)} = \frac{\delta +
\left(1 - \delta \right) y_0^m - \delta \left( 1 - y_0 \right)^n}
{\delta + \left(1 - \delta \right) y_1^m - \delta \left( 1 - y_1
\right)^n}~,
\ee
where $y_0 = y(t_0)$ and $y_1 = y(t_1)$.
We then have a standard null geodesic equation
\be
\label{geod}
\int_0^{r_1} \frac{dr}{\sqrt{1-kr^2}} = \int_{t_1}^{t_0}
\frac{cdt}{a(t)}~,
\ee
with the scale factor $a(t)$ given by (\ref{sf2}). For a flat Friedmann model we can write down the radial coordinate to an observer in any of the forms
\bea
\label{radial}
r_1 &=& \int_{t_1}^{t_0} \frac{cdt}{a(t)} = \int_{y_1}^{y_0} \frac{(ct_s) dy}{a(y)} = \int_{t_1}^{t_0}
\frac{cda}{\dot{a}a} = \int_{a_1}^{a_0} \frac{cda}{Ha^2} \nonumber
\\ &=& \int_{0}^{z} \frac{cdz}{H(z)a_0} = \frac{c}{H_0a_0} \int_{0}^{z}
\frac{dz}{E(z)}~,
\eea
where we have used that $y=t/t_s$ and so $dy = dt/t_s$ which implies $a'(y) \equiv da(y)/dy = t_s da(t)/dt = t_s \dot{a}(t)$, $a'(y_0) = t_s \dot{a}(t_0)$ and
\be
\label{Hy}
H_0 \equiv H(t_0) =  \frac{1}{t_s} \frac{a^{\prime}(y_0)}{a_0} = \frac{1}{t_s} H(y_0)~\hspace{0.2cm}; \hspace{0.2cm} H(y) \equiv \frac{a'(y)}{a(y)}~.
\ee
In (\ref{radial}) the transition from the integral of $da$ to the integral of $dz$ was given by the application of the definition of redshift (\ref{redshift}). Besides, due a lack of an analytic form for the equation of state for SFS models the function $E(z)$ can only be given by a formula which involves an integral over $z$, as follows
\be
\label{E(z)}
E^2(z) = \frac{H^2(z)}{H_0^2} = \Omega_{\delta 0} (1+z)^3 \exp{\left[\int_{0}^{z}
dz' \frac{2q(z') - 1}{1+z'} \right]}~,
\ee
in (\ref{radial}) with $\Omega_{\delta 0}$ being the density parameter of sudden-future-singularity-driven dark energy \cite{aps10}. It is easy to notice that in the limit $\delta \to 0$ and $m=2/3$ one has that $E(z) = \Omega_{\delta 0} (1+z)^3$, as for the standard matter dominated case. Let us recall that the standard formula for the models which include the dark energy component $\Omega_{w0}$ reads as \cite{PRD41,Nesseris:2006er}
\be
\label{standEz}
E^2(z) = \Omega_{m0} (1+z)^3 + \Omega_{w0} (1+z)^{3(w+1)} + \Omega_{k0}(1+z)^2~,
\ee
where $\Omega_{k0}$ is the curvature component and $p = w\varrho$. However, in our further calculations we will not be expressing $E(z)$ in either forms (\ref{E(z)}) or (\ref{standEz}), using an explicit form of $a(y)$ as in (\ref{radial}) instead.

The luminosity distance to e.g. a supernova observed at redshift, $z$, is given by
\be
d_L(z)=(1+z)a(t_0)r_1
\ee
where $r_1$ is given in one of the forms (\ref{radial}). For our calculations we have applied the following expression
\be
d_{L}(z) = (1+z)a(y_0) c t_s \int_{y_1}^{y_0}
\frac{dy}{a(y)}~.
\ee
The distance modulus is
\be
\mu(z)=5\log_{10}d_L(z)+25.
\ee
We compared the luminosity distance predicted for SFS models with SNIa data using the SCP Union2 \cite{Amanullah} dataset, consisting of 557 supernovae, which is the largest compilation published to date.

\section{Shift parameter}
\setcounter{equation}{0}
\label{shift}

The standard formula for the CMB shift parameter is given by
\cite{bond97,Nesseris:2006er}:
\be
\label{shifteq}
{\cal R}=\frac{l_1^{\prime TT}}{l_1^{TT}}~,
\ee
where $l_1^{TT}$ is the temperature perturbation CMB spectrum multipole of the first acoustic peak in the model under consideration and $l_1^{\prime TT}$ corresponds to a reference flat standard Cold Dark Matter (CDM) model. The multipole number is related to an angular scale of the sound horizon $r_s$ at decoupling by \cite{PRD41,singh03}
\be
\label{theta}
\theta_1 = \frac{r_s}{d_A} \propto \frac{1}{l_1}~,
\ee
where
\be
r_s = a_{\rm dec} S(r_s) = a_{\rm dec} S\left( \int_{0}^{t_{\rm dec}} c_s \frac{dt}{a(t)} \right)~,
\ee
with $c_s$ being the sound velocity and the angular diameter distance reads as
\be
d_A = a_{\rm dec} S(r_{\rm dec}) = a_{\rm dec} S\left( \int_{t_{\rm dec}}^{t_0} c \frac{dt}{a(t)} \right)
\ee
where $c$ is the velocity of light and $S(r) = r$ for $k=0$.

Following \cite{Nesseris:2006er} we can write, using \ref{shifteq} and \ref{theta} as follows
\be
{\cal R} = \frac{r_s}{r'_s} \frac{d'_A(z'_{\rm dec})}{d_A(z_{\rm dec})}~,
\ee
which, by assuming that at decoupling the amount of radiation was the same in both the flat reference standard CDM model and in our SFS model (which we assume to be just the same as a standard matter-radiation model of the early universe, since SFS models do not change the evolution there) we have that
\be
\frac{r_s}{r'_s} = \frac{1}{\sqrt{\Omega_{m0}}}~.
\ee
On the other hand, for a reference standard CDM model
\bea
d'_A &=& \frac{2c a'_{\rm dec}}{a_0H_0} \left[\sqrt{\Omega'_{r0} + 1}
- \sqrt{\Omega'_{r0} + \frac{a'_{\rm dec}}{a_0}} \right] \nonumber \\
&=& \frac{2c a'_{\rm dec}}{a_0H_0} f(\Omega'_{r0},a'_{\rm dec})~,
\eea
while for our SFS model the angular diameter distance is given by
\be
d_A = a_{\rm dec} r_{\rm dec}
\ee
with $r_{\rm dec}$ given by (\ref{radial}) taken at decoupling.
Using the above, we may write that for our SFS models the shift parameter is
\be
{\cal R} =
\frac{2c}{H_0a_0 \sqrt{\Omega_{m0}} r_{\rm dec}} = \frac{2ct_s}{a'(y_0) \sqrt{\Omega_{m0}} r_{\rm dec}}~,
\ee
where we have assumed that the function $f(\Omega'_{r0},a'_{\rm dec})$ is approximately unity \cite{Nesseris:2006er}.

Finally, the rescaled shift parameter is
\bea
{\cal \bar{R}}&=&\sqrt{\Omega_{m0}} \frac{H_0a_0}{c} r_{\rm dec} = \sqrt{\Omega_{m0}} \int_0^z\frac{dz}{E(z)} \nonumber \\
&=& \sqrt{\Omega_{m0}} a'(y_0) \int_{y_{\rm dec}}^{y_0} \frac{dy}{a(y)},
\eea
where in order to obtain the last expression we have used (\ref{Hy}). The WMAP data gives ${\cal R}=1.70\pm0.03$ \cite{wang}.
Note that, as we discussed in \cite{ghodsi11}, this value is not directly ``observed'' but is derived from the CMB data assuming a specific class of cosmological model; hence one must be careful in employing it to test our SFS model. However, as  we showed in \cite{ghodsi11}, the effective equation of state for our SFS model displays general similarity to that of the concordance model and on this basis we consider the use of the ``observed" value of the shift parameter to be appropriate for our SFS model too.

Moreover in \cite{ghodsi11} we also considered the acoustic scale, $l_a$, and how its incorporation into the analysis could result in parameter constraints which gave a closer approximation to those obtained from a fit to the full CMB power spectrum. However,  upon varying the parameter $m$ from the Einstein-de Sitter value of $2/3$ considered in \cite{ghodsi11}, it is not clear that the use of the acoustic scale remains valid.  To see this we recall that the acoustic scale is defined in terms of two distinct distance scales: it is proportional to the ratio of the angular diameter distance, $d_A$, and the sound horizon, $r_s$, both evaluated at the redshift of recombination, $z_{CMB}$:
\bea
l_a=\pi\frac{d_A(z_{CMB})}{r_s(z_{CMB})},
\eea
where the comoving sound horizon is defined as:
\bea
r_s=\int_{z_{CMB}}^\infty \frac{c_sdz'}{E(z')} \quad \textrm{with} \quad c_s=\frac{1}{\sqrt{3(1+\overline R_b a)}},
\eea
where $E(z)=H(z)/H_0$ and $c_s$ is the speed of sound with $\overline R_b=31500 \Omega_b h^2 (\frac{T_{CMB}}{2.7k})^{-4}, T_{CMB} = 2.7$K. Now the potential difficulty in carrying out the $l_a$ calculation for the $m \neq 2/3$ case arises from the sound speed calculation. Previously in the $m=2/3$ case we could interpret the sound speed as being set at the time of recombination in the ``standard early universe''. The value of the acoustic scale would still in general be different for our SFS universe due to different expansion history of the model and hence different angular diameter distance-redshift relation.  In the case of $m \neq 2/3$, however, the above simplification is no longer valid and {\em both\/} the sound horizon and angular diameter distance will in general be different for our SFS universe.  In particular computing the sound speed in the $m \neq 2/3$ case would require more detailed modelling of the SFS fluid in the early universe which lies beyond the scope of this paper.  Hence we do not consider the acoustic scale further in this paper.

\section{Baryon acoustic oscillations}
\setcounter{equation}{0}
\label{BAO}

The Alcock-Paczy\'nski effect \cite{AP} states that one is able to calculate the distortion of a spherical object in the sky without knowing its true size. This can be done by measuring its transverse extent using the angular diameter distance, $r$
\be
\label{r}
r = \frac{l}{\Delta \theta}~,
\ee
where $l$ and $\Delta \theta$ are the linear and angular size of an object, and its line-of-sight extent, $\Delta r$, using the redshift distance
\be
\label{dr}
\Delta r = \frac{c \Delta t}{a(t)} = \frac{c t_s \Delta y}{a(y)}
\ee
(see e.g. Ref. \cite{Nesseris:2006er}). As a result one can define the volume distance, $D_V$, as
\be
D_V^3 = r^2 \Delta r~~,
\ee
so that using (\ref{radial}), one has
\bea
\label{DV2}
D_V &=& \left[\left(\int_{y_1}^{y_0} \frac{ct_sdy}{a(y)} \right)^2
\left( \frac{ct_s \Delta y}{a(y)} \right) \right]^{\frac{1}{3}} \\
\label{DV3}
&=& \left[\left(\frac{c}{a_0H_0} \int_0^z \frac{dz}{E(z)} \right)^2 \left(\frac{c}{a_0H_0} \frac{\Delta z}{E(z)} \right) \right]^{\frac{1}{3}}~~.
\eea
Ref. \cite{eisenstein} uses the acoustic peak scale for 46748 luminous red galaxies selected from the Sloan Digital Sky Survey to measure $D_V (\Delta z = z_{\rm BAO}=0.35) = 1370 \pm 64$ Mpc.

Usually, it is more convenient to work with a dimensionless quantity ${\cal A}$ which for our SFS model (\ref{sf2}) is obtained multiplying (\ref{DV3}) by $\sqrt{\Omega_{m0}} (a_0 H_0)/(c z_{\rm BAO})$ to obtain
\be
{\cal A}= \Omega_{0m}^{1/2}E(z_{\rm BAO})^{-1/3}\left[\frac{1}{z_{\rm BAO}}
\int^{z_1}_0\frac{dz}{E(z)} \right]^{2/3}~,
\ee
or, using the definition of $E(z)$ given by the first equality in (\ref{E(z)}), together with (\ref{Hy}),
\bea
{\cal A}= \frac{\sqrt{\Omega_{m0}} a^{\prime}(y_0)}{[a^{\prime}(y_{\rm BAO})]^{\frac{1}{3}}}
\left[\frac{a(y_{\rm BAO})}
{a(y_0)}\right]^{\frac{1}{3}}
\left[\frac{1}{z_{\rm BAO}}\int_{y_{\rm BAO}}^{y_0}\frac{dy}{a(y)}\right]^{\frac{2}{3}}
\eea
The same result can be obtained from (\ref{DV2}) if we compute the difference $\Delta y = y_0 - y_{\rm BAO}$ and $a(y) = a(y_0)$ in (\ref{dr}), i.e.,
\be
\label{DV4}
D_V = ct_s \left[ \frac{y_0 - y_{\rm BAO}}{a(y_0)}
\left(\int_{y_{\rm BAO}}^{y_0} \frac{dy}{a(y)} \right)^2
\right]^{\frac{1}{3}}
\ee
and use the fact that $ct_s = c/H_0 a'(y_0)/a(y_0)$.
Following Ref. \cite{eisenstein} the parameter ${\cal A}$
should have the value
\be
{\cal A} = 0.469 \left( \frac{n}{0.98} \right)^{-0.35} \pm 0.017~~,
\ee
where $n$ is the spectral index (now taken to be $\sim 0.96$).

\section{Results and conclusions}
\setcounter{equation}{0}
\label{conclusion}

We used a Bayesian framework to confront our SFS model with the cosmological observations discussed in the previous sections. For each cosmological probe we took the likelihood function to be Gaussian in form, i.e.
\be
p({\rm data} | \Theta) \propto \exp ( - \frac{1}{2} \chi^2),
\ee
where $\Theta$ denotes the parameters of the SFS model and ``data" denotes generically the observed data for one of the three cosmological probes.  For the SNIa data $\chi^2$ takes the form
\be
\chi^2_{\rm SN}=\sum^{N}_{i=1}\frac{(\mu_{\rm obs}(z_i)-\mu_{\rm pred}(z_i))^2}{\sigma^2_i+\sigma_{\rm int}^2}~,
\ee
where $\sigma_i$ is the quoted observational error on the $i^{\rm th}$ Union2 SNIa and $\sigma_{\rm int}$ is the SNIa intrinsic scatter. Following
\cite{Amanullah} we take $\sigma_{\rm int}=0.15$. For the CMB shift parameter $\chi^2$ takes the form
\be
\chi^2_{\cal R}=\frac{({\cal R}-1.70)^2}{0.03^2}~,
\ee
while for the BAO distance parameter it takes the form
\be
\chi^2_{\cal A}=\frac{({\cal A}-0.469)^2}{0.017^2}~.
\ee
Since our three cosmological probes are mutually independent, their joint likelihood function is given by the product of their individual likelihoods, and thus takes the form
\be
p({\rm all \, \, data} | \Theta) \propto \exp ( - \frac{1}{2} \chi^2_{\rm TOT}),
\ee
where $\chi^2_{\rm TOT} = \chi^2_{\rm SN} + \chi^2_{\cal R} + \chi^2_{\cal A}$.

We used Bayes' theorem and a Markov Chain Monte Carlo (MCMC) approach \cite{philgregory} to obtain posterior probability distributions for the SFS model parameters: $n$, $m$, $\delta$ and $y_0={t_0}/{t_s}$, where $t_0$ is the present age of the universe. We employed the Metropolis-Hastings algorithm, adopting uniform priors for each of the parameters: $\delta\in(-30,0)$, $n\in(1,2)$, $m\in(0,3)$, $y_0\in(0,1)$.

\begin{figure*}
    \begin{tabular}{ccc}
      \resizebox{50mm}{!}{\rotatebox{-90}{\includegraphics{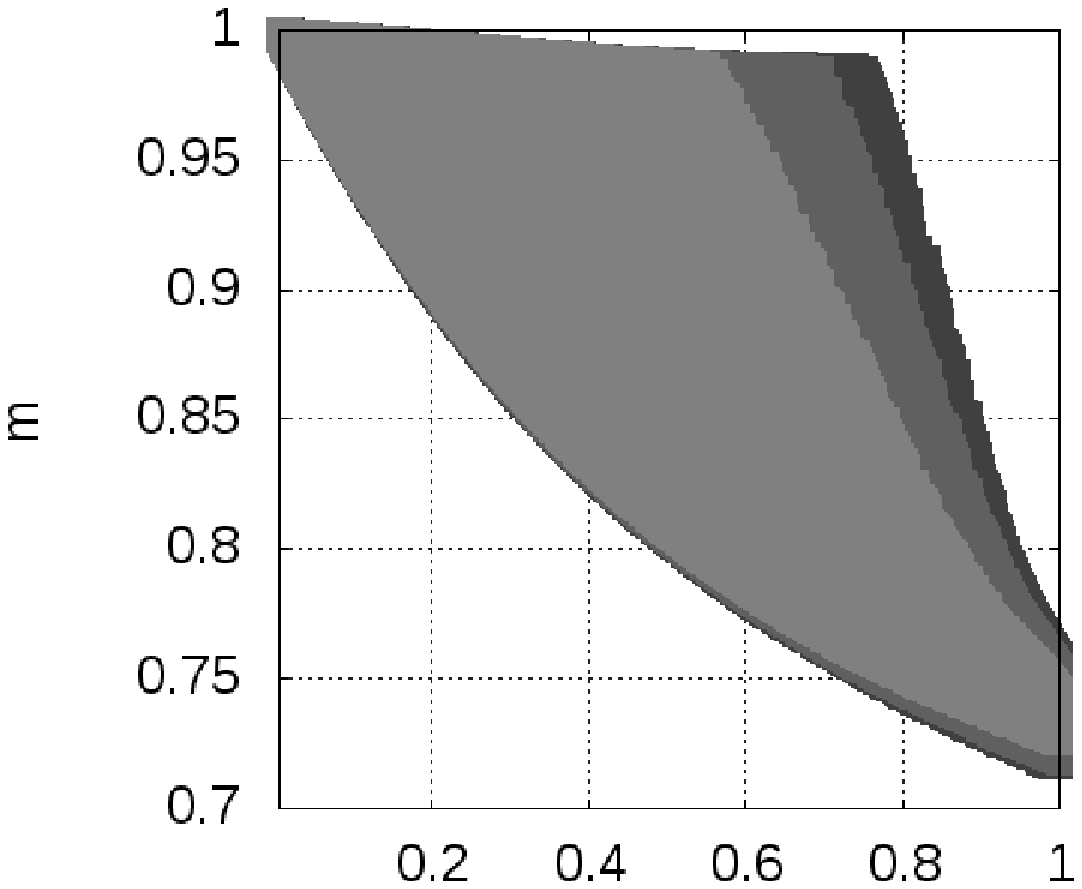}}} &
       &
       \\

      \resizebox{50mm}{!}{\rotatebox{-90}{\includegraphics{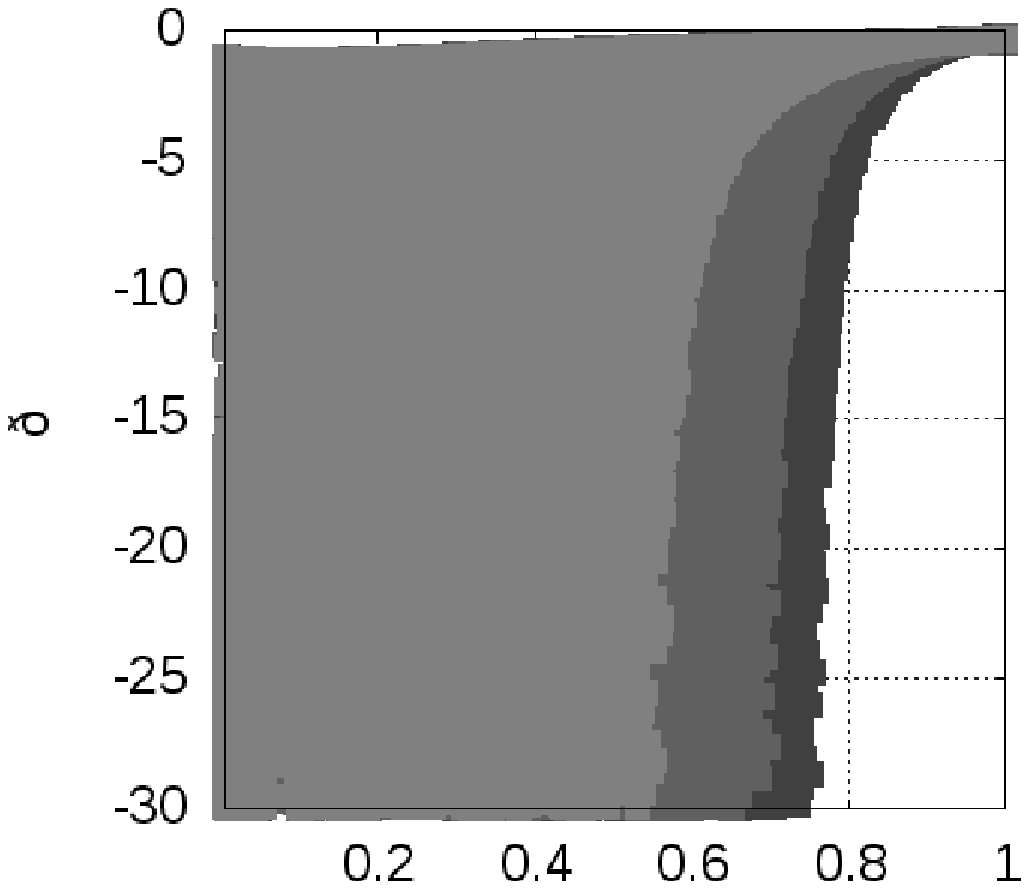}}} &
      \resizebox{50mm}{!}{\rotatebox{-90}{\includegraphics{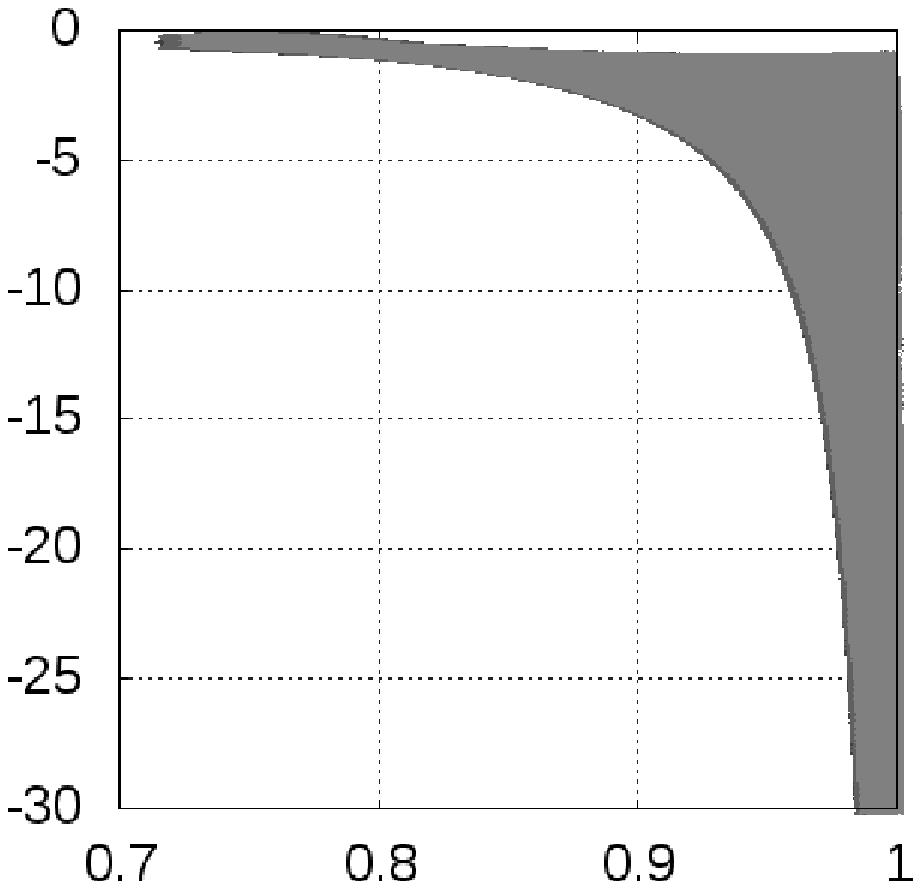}}} &
       \\

      \resizebox{50mm}{!}{\rotatebox{-90}{\includegraphics{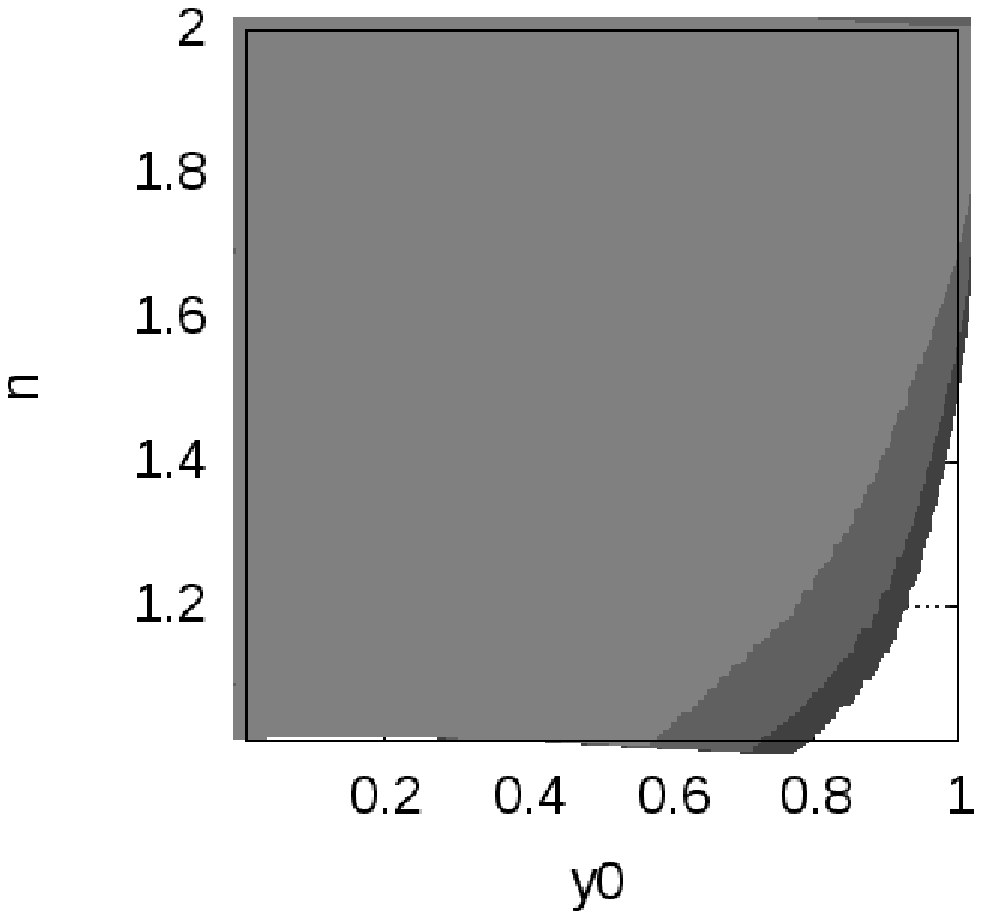}}} &
      \resizebox{50mm}{!}{\rotatebox{-90}{\includegraphics{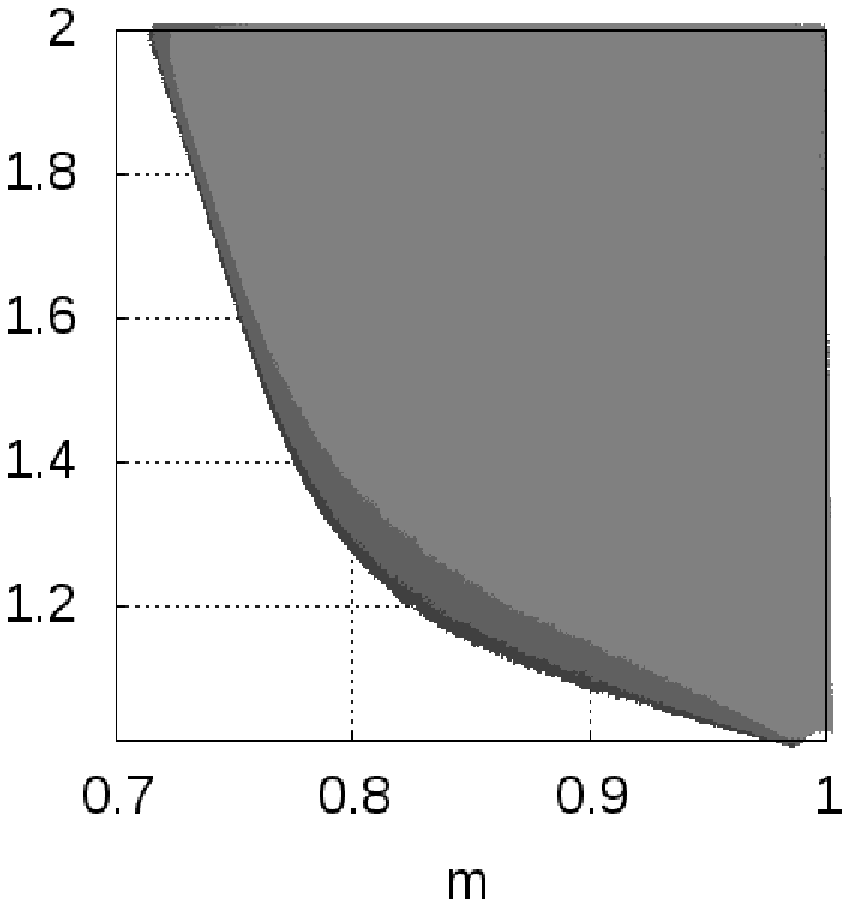}}} &
      \resizebox{50mm}{!}{\rotatebox{-90}{\includegraphics{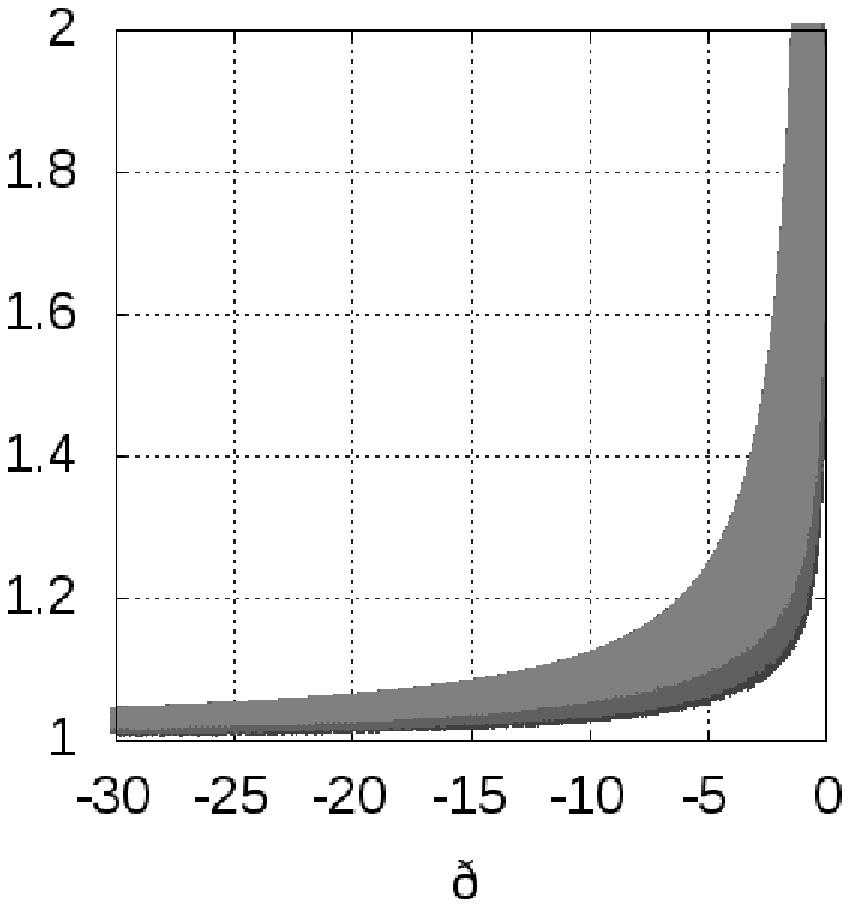}}} \\

    \end{tabular}

\caption{Marginalized contours for pairs of parameters are plotted. There are three levels of credible region shown: $68.3\%$, $95.4\%$, $99.73\%$ (from light grey to dark grey respectively) calculated from $\cal{A},\ \cal{R},$ and SN Ia jointly}\label{fig1}
\end{figure*}

In Fig. \ref{fig1} we present contour plots showing the joint marginal posterior distribution for each pair of SFS model parameters.  Each sub-panel shows three contours, denoting $68.3\%$, $95.4\%$ and $99.73\%$ (from light gray to dark grey respectively) credible regions.

In Fig. \ref{fig2} we also show marginal posterior distributions for each of the model parameters individually, again determined via Markov Chain Monte Carlo.
It is immediately clear from Figs. \ref{fig1} and \ref{fig2} that the current cosmological data strongly exclude our SFS model for
$m = 2/3$ (indicated by the dashed horizontal and vertical lines in the relevant sub-panels of each figure).  In the leftmost panel of Fig. \ref{fig2}, for example, we see that the marginal posterior distribution for $m$ is strongly peaked between 0.7 and 1.0, but is negligible outside the range $(0.7,1.0)$.  In other words our SFS model is incompatible with current observations in the particular case where the asymptotic behaviour of the scale factor close to the BB singularity mimics a dust-filled Einstein de Sitter universe.  This of course is not surprising since as noted in Section \ref{intro} it was the conclusion we reached in \cite{ghodsi11}.
\begin{figure*}
\includegraphics[width=15cm,height=4.5cm,angle=0]{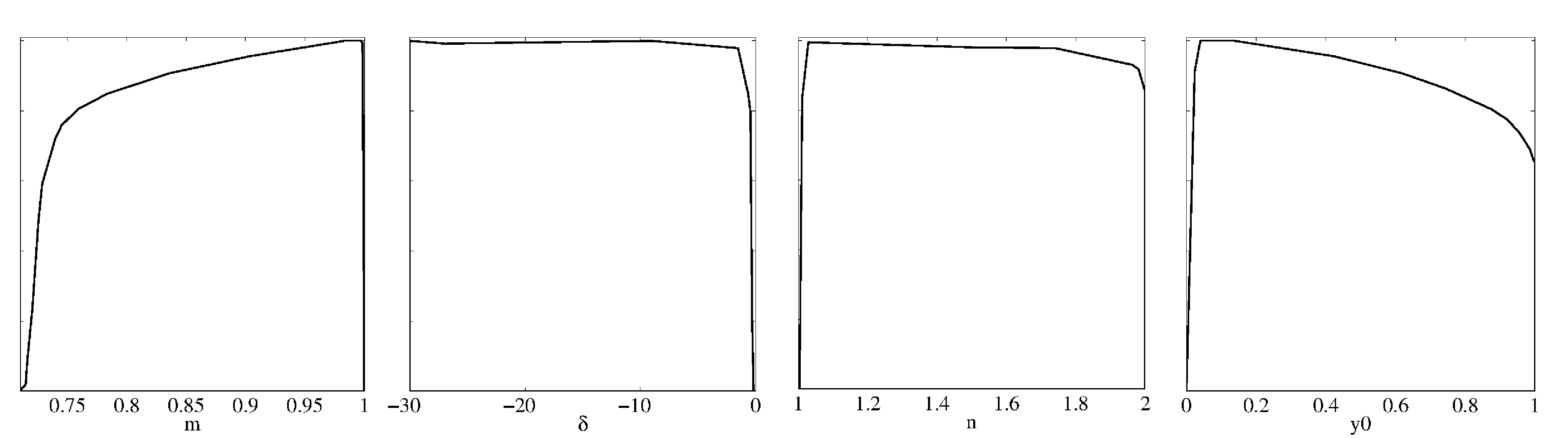}
\caption{Unnormalized probability densities for all parameters, calculated for $\cal{A},\ \cal{R},$ and SN Ia jointly.} \label{fig2}
\end{figure*}
On the other hand it is also clear from Figs. \ref{fig1} and \ref{fig2} that there is a significant region of parameter space where the marginal posterior distributions of the SFS model parameters appear to overlap with each other -- i.e. one can identify sets of model parameters which occupy the $68.3\%$ credible region for {\em all\/} of the marginal posterior distributions shown.  This does not in itself guarantee that these sets of model parameters give an acceptable fit to the SFS model since the marginal posterior distributions are projections of the fully, four-dimensional joint posterior: the apparent overlap of the two-dimensional marginal distributions for each pair of parameters, for example, could in principle be merely a projection effect.  We have checked this rigorously, however, via the following approach.  We considered each of the three cosmological probes separately and produced MCMC chains that explored the full four-dimensional SFS model parameter space for each probe. We then identified the MCMC chain points that sampled e.g. the $68.3\%$ credible region for the SNIa chain and verified (via direct computation of the likelihood values) that there was significant overlap in {\em four\/} dimensions between these points and the corresponding credible regions of both other probes.  Thus we established that the apparent overlap of the marginal posterior distributions for the joint likelihood of all three cosmological probes was not merely a projection effect.

It is clear from this analysis, therefore, that the SFS model under consideration {\em is\/} compatible with current cosmological data, based on the three cosmological probes that we considered.  In particular, provided that $m \geq 0.72$ we see that we can identify ranges for the other parameters for which the predicted values of the cosmological probes are in good agreement with current observations.  As the value of $m$ increases towards unity the credible regions for the parameters $\delta$ and $n$ extend to fill their allowed prior ranges, while the credible region for $y_0$ is pushed towards lower values -- i.e. the SFS model is still compatible with current data but the epoch of the sudden future singularity is pushed further into the future.  Interestingly, however, we see that as $m \rightarrow 0.72$ the credible regions for $n$ and $\delta$ become much narrower, constrained to lie ever closer to $n=2$ and $\delta = 0$, while the credible region for $y_0$ tends towards unity, so that the SFS may happen in the very near future.

Thus we conclude that our SFS models are compatible with current observations provided the parameter $m$, which characterizes the near-to-big-bang evolution of the scale factor, is at least $0.72$. This excludes the Einstein de Sitter dust solution, and requires a form of matter which has slightly negative pressure (since $m=2/3(w+1)$, so that $w \leq -0.083 $). Note also that, in order to match the current observations, the value of the ``non-standardicity" parameter $\delta$ is seen necessarily to be negative.  This naturally eliminates the validity of anti-Chaplygin gas models \cite{kamenschchik} given in a special form by (\ref{achap2}) which require $\delta \to 1$ (but it does not eliminate the models in the form of (\ref{achap1}) despite their static Friedmann limit and also other types of anti-Chaplygin gas models studied in Ref. \cite{laszlo}).

In conclusion, similarly as in our previous paper \cite{PRD07} where we applied supernovae data only, here we have shown that a sudden future singularity may happen in the near future of the universe.

\section{Acknowledgements}

T.D. and M.P.D. acknowledge the support of the National Science Center grant No N N202 3269 40. We thank John Barrow, Leonardo Fernandez-Jambrina, Laszlo Gergely and Sergei Odintsov for discussions.\\
\indent H.G. would like to acknowledge the fact that she performed the majority of her part of this work while at the University of Glasgow.\\
\indent Part of the simulations reported in this work were performed using the HPC cluster HAL9000 of the Computing Centre of the Faculty of Mathematics and Physics at the University of Szczecin.

\end{document}